# Features distinguishing the flow behavior of polyelectrolytes with opposite charges in aqueous solutions

Suresha P. Ranganath, Manohar V. Badiger, and Bernhard A. Wolf *

**ABSTRACT**

Solution viscosities of poly(3 acryl amido propyl trimethyl ammonium chloride), (PAPTMAC-Cl 7.8) and of poly(styrene sulfonate sodium) (PSS-Na 75.6) were measured in water containing different amounts of NaCl at 25 °C; the numbers after the abbreviation of the polymers state their $M_w$ in kDa. The intrinsic viscosity of the polycation in water amounts to 4460 mL/g and is about three times larger than that of the polyanion, despite the fact that its mass is only one tenth that of the polycation. At low salinities the shear-overlap parameters $\Sigma$ as a function of composition $c$ (mass per volume) pass maxima for PAPTMAC-Cl 7.8, but exhibit points of inflection for PSS-Na 75.6. At sufficiently large salinities these dependencies become linear in both cases. Points of inflection in $\Sigma(c)$ indicate cross-over condition and separate the solvent dominated from the solute dominated flow regimes. For the polycation, the $\Sigma_{crov}$ values are always larger than for the polyanion. The analysis of these results in the light of the HSAB (Hard Soft Acid Base) approach of Pearson yields strong indications that the observed dissimilarities between the two types of polyelectrolytes are not caused by the opposite charges attached to the macromolecules, but result from the combination of the *soft* $NR_4^+$ with the *hard* $Cl^-$ ions in the case of PAPTMAC-Cl 7.8 and of the *hard* $RSO_3^-$ with the *hard* $Na^+$ ions in the case of PSS-Na 75.6. This insight opens up new opportunities for the tailoring of rheological properties for polyelectrolytes in saline solutions.

## 1. Introduction

This study was initiated with the objective to determine whether there exist fundamental differences in the flow behavior of polyanion and polycation solutions. However, it turned out that the question was wrongly posed, since the observations reported here cannot be rationalized by the opposite sign of the charges attached to the macromolecules. Consequently, we have searched recent review articles[1-4] to see whether they provide explanations for the unexpected findings, but in vain. We have therefore developed our own ideas and hypotheses as shown in this contribution. They are based on a number of earlier observations and tools, which we briefly outline below to make the presentation easier to understand.

The substitution of Huggins plots or similar extrapolation methods for the determination of intrinsic viscosities by an alternative procedure is probably the most important one, because it enables the flawless measurement in the case of polyelectrolytes.[5] Plotting the natural logarithm of the relative viscosity ($\eta_{rel} = \eta_{solution}/\eta_{solvent}$) versus the polymer concentration $c$ yields unadulterated $[\eta]$ values and avoids the artifacts of maxima caused by the zero divided by situation of the extrapolation of ($\eta_{rel}$ -1)/$c$ to infinite dilution. In all the cases studied so far, the viscosities of the polymer solution increase steadily with rising polymer concentration and lack maxima. Another important item concerns the possibility to describe ln $\eta_{rel}$ as a function of $c$ quantitative; a simple expression with two adjustable parameters models this dependence over the full range of composition[6] in the case of simple systems; for more complex systems, like the solutions of star-like polyelectrolytes, a third parameter turns out to be necessary.[7]

The extension of the concept of intrinsic viscosities to **generalized** intrinsic viscosities[8], $\{\eta\}$, constitutes another important tool for the present investigation; $\{\eta\}$ represents the hydrodynamic specific volume (volume/mass) at arbitrary polymer concentration but not at infinite dilution - like $[\eta]$. Furthermore, the introduction of the generalized intrinsic viscosity is a prerequisite for the study of the shear overlap parameter $\Sigma$ and of the cross-over conditions.[9] $\Sigma$ vs $c$ indicates how many solute molecules flow together under given conditions at different polymer concentrations and the cross-over concentration separates the solvent dominated flow regime (*less* than linear increase of $\Sigma$ with $c$) from the polymer dominated flow regime (*more* than linear increase of $\Sigma$ with $c$).[9]

As already mentioned at the beginning, the results reported here cannot simply be explained by the opposite signs of the charges attached to the polymer. This leads to the conclusion that substance-specific interactions between them and their low molecular weight counterions could cause the behavior described here. In fact, such ideas have been discussed as early as 1888 in the field of inorganic chemistry[10] (Hofmeister series). More recent approaches group the different ions into chaotropic and kosmotropic[11] or into hard or soft acids or bases.[12, 13] In view of the fact that the HSAB (Hard Soft Acid Base) principle of Pearson has proven to be an effective explanatory tool in the field of inorganic chemistry, as evidenced by over 1,400 citations in the Web of Science, we have decided to discuss the present findings on the basis of this approach.

## 2. Theoretical background

This paper is based on the quantitative description of the viscosity $\eta$ of polymer solutions as a function of the polymer concentration $c$ (mass/volume). The probably most used relation for this purpose is the Huggins equation, a series expansion in terms of the reduced polymer concentration $\tilde{c}$

$$\eta = \eta_o(1 + \tilde{c} + k_H \tilde{c}^2 + \cdots) \qquad (1)$$

where $\tilde{c}$ is defined as the product of $c$ and the intrinsic viscosity $[\eta]$

$$\tilde{c} = c[\eta] \qquad (2)$$

The reason, why the Huggins relation is so frequently used lies in its possibility to obtain quick and easy access to information on the molar mass of the polymers via $[\eta]$ vs $M$ relationships. For uncharged polymers eq (1) works well; for polyelectrolytes, however, it fails because of a zero divided by zero situation upon extrapolation to infinite dilution.[6]

To cover larger concentrations intervals, scaling laws have so far predominantly been used, [14-16] where a minimum of two scaling parameters is normally needed: The correlation length (or blob size) and the entanglement concentration, marking the crossover from a dilute to a semi-dilute or concentrated regime. This approach offers a simple and intuitive way to understand the universal features of polymer solutions, without needing complex microscopic models. In practice, double logarithmic plots are commonly used to find out linear ranges; these can for instance be $(\eta - \eta)_o/\eta_o$ vs $n_P$, the number of monomers per volume[14, 15] or $(\eta - \eta)_o/\eta_o$ vs $c$ as in references[16].

The present approach rests on the introduction of a generalized intrinsic viscosity[8] $\{\eta\}$ as

$$\{\eta\} = \left(\frac{\partial \ln \eta}{\partial c}\right) = \left(\frac{\partial \ln \eta_{\text{rel}}}{\partial c}\right). \qquad (3)$$

This quantity represents the hydrodynamic specific volumed of the polymer *at arbitrary concentration c* (mass/volume) in contrast to the intrinsic viscosity $[\eta]$ which refers to infinite dilution.

$$[\eta] = \lim_{c \to 0} \left( \frac{\partial \ln \eta_{rel}}{\partial c} \right) \qquad (4)$$

Similar to the limit of the $\{\eta\}$ at infinite dilution, there is also a corresponding limit for vanishing solvent content: the intrinsic bulkiness $⟦\eta⟧$, as formulated below

$$⟦\eta⟧ = \lim_{c \to \rho_{pol}} \left( \frac{\partial \ln \eta_{rel}}{\partial c} \right) \qquad (5)$$

Figure 3 illustrates this situation graphically.

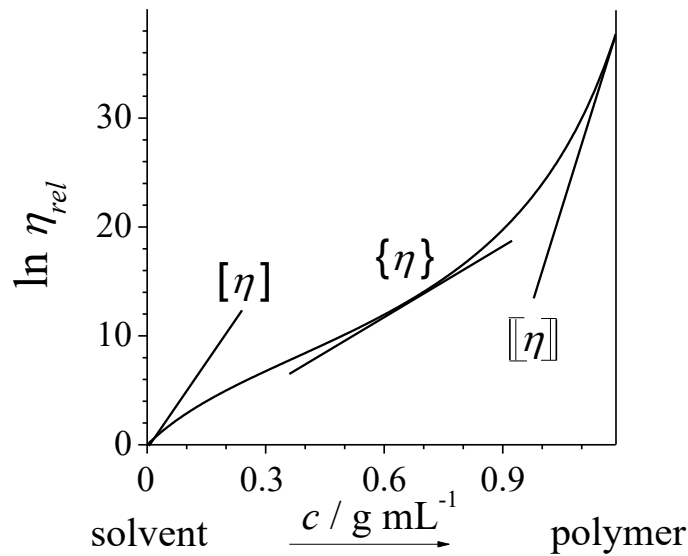

**Figure 1**: Scheme showing how the system specific parameters $[\eta]$, $\{\eta\}$ and $⟦\eta⟧$ are determined from the concentration dependence of the relative viscosity $\eta_{rel}$.

The curve shown in Fig. 1 applies to temperatures above the glass transition temperature of the pure polymer. If $T$ is lower, ln $\eta_{rel}$ approaches infinity at the glass transition concentration, where the solution loses its liquid structure[6]; under these circumstances, it is no longer possible to obtain physically meaningful values for the intrinsic bulkiness.

The generalized intrinsic viscosity can either be obtained by determining the slope of ln $\eta_{rel}$ vs $c$ graphically or through mathematical modeling of the dependence by means of the obtained parameters. The following expression proved to be the best for that purpose.

$$\ln \eta = \ln \eta_o + \frac{\tilde{c} + \alpha \tilde{c}^2}{1 + \beta \tilde{c} + \gamma \tilde{c}^2} \qquad (6)$$

where $\alpha$, $\beta$ and $\gamma$ are hydrodynamic interaction parameters with a clear-cut physical meaning, which becomes obvious if the above relation is rewritten in form of a series expansion in terms of the reduced polymer concentration $\tilde{c}$ as

$$\ln \eta = \ln \eta_o + \tilde{c} + (\alpha - \beta)\tilde{c}^2 + (\beta^2 - \alpha\beta - \gamma)\tilde{c}^3 + \cdots \qquad (7)$$

The second term, $(\alpha - \beta)$, accounts for the friction between a solvent molecule and *two* segments belonging to *different* macromolecules and the third term, $(\beta^2 - \alpha\beta - \gamma)$, to the friction between a solvent molecule and *three* such intermolecular contacts.

The application of eq (6) to a large number of polymer solution has demonstrated[6, 7, 17-22] that only two of them (normally $\beta$ and $\alpha$ or $\gamma$) are in the majority of cases required to model the entire concentrations range of interest. The typical experimental errors of the viscometric interaction parameters lie on the order of ± 2 to 4 % for $\beta$; that of $\alpha$ and $\gamma$ (the perfectors of $\tilde{c}^2$) may become twice as high.

The combination of eqs (3) and (6) yields the following expression,[8] which we used throughout the paper to calculate the generalized intrinsic viscosities by means of the viscometric interaction parameters.

$$\{\eta\} = \frac{[\eta]\left(1 + 2\alpha\tilde{c} + (\alpha\beta - \gamma)\tilde{c}^2\right)}{\left(1 + \beta\tilde{c} + \gamma\tilde{c}^2\right)^2} \qquad (8)$$

In order to estimate the uncertainty in $\{\eta\}$, resulting from that of the viscometric interaction parameters, we performed a Gaussian error analysis and determined the probable absolute error as a function of polymer concentration in the range of interest. For easier comparison we have then converted these data into relative errors and obtained the following results: In the absence of extra salt and PAPTMAC-Cl 7.8 the error starts from 4,7% at high dilution and increases up to approximately 10% for the highest polymer concentrations; for PSS-Na 75.6 the error remains on the order of 5-6 % within the entire concentration range. As the salinity of the solvent rises, the errors decrease rapidly in both cases.

The introduction of the generalized intrinsic viscosity as the central parameter of the present approach brings with it another major advantage, namely the possibility to calculate the size of molecular aggregates formed during viscous flow. This is so, because $\{\eta\}$ is related to $c_{\text{shear cluster}}$, the concentration of an *individual* solute molecule within the shear cluster[9, 23] by

$$c_{\text{shear cluster}} = \frac{1}{\{\eta\}} \qquad (9)$$

and yields the shear overlap parameter $\Sigma$ - stating the number of solute molecules that a selected solute molecule is flowing with - according to the following simple relationship[9, 23]

$$\Sigma = \frac{c}{c_{\text{shear cluster}}} = c\,\{\eta\} \qquad (10)$$

c = 0 yields $\Sigma$ = 0, meaning that each macromolecule flows in isolation. Therefore, the number of polymer molecules in a cluster is equal to $\Sigma + 1$.

The concentration dependence of $\Sigma$ provides information on the so-called cross-over concentrations[9, 23], $c_{\text{cross-over}}$. This parameter subdivides the concentration regime into a dilute region - in which the solvent dominates the flow behavior - and a concentrated region - in which the polymer dominates. The condition that needs to be fulfilled at this characteristic point is as follows:

$$\frac{\partial^2 \Sigma}{\partial c^2} = 0 \qquad (11)$$

Figure 1 shows for a typical aqueous polyelectrolyte solution in the absence of extra salt how the size of shear clusters varies with concentration. For such systems, the crossover conditions are already reached at extremely low $c$ values of around 0.0001 g/mL due to the large spatial extension of the polymer coils in the limit of infinite dilution. For the system on which the sketch is based, the volume of individual coils shrinks to approximately one-twentieth as $c$ increases, due to self-shielding.[19]

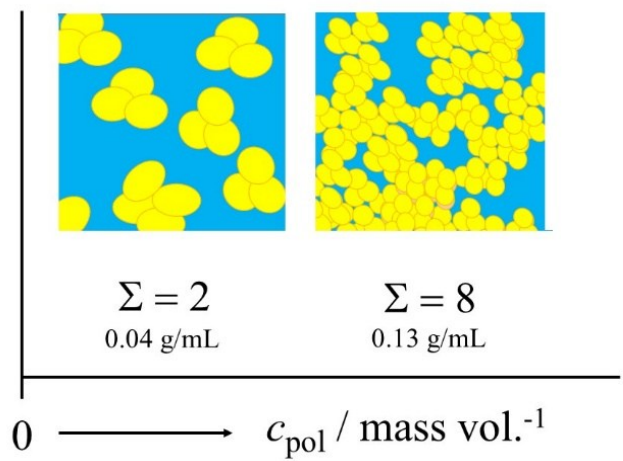

**Figure 2**: Sketch of the (average) number $\Sigma$ of polymer molecules flowing together at different polymer concentrations, based on the experimental results for PSS-Na 75.6 solution in pure water.

3. **Materials and procedures**

**PAPTMAC-Cl 7.8 -** (poly(3 acryl amido propyl trimethyl ammonium chloride)) - the polycation under investigation was synthesized and characterized in the following manner. Monomer, 3 acrylamidopropyl trimethyl ammonium chloride (APTMAC) (75 wt-% aqueous solution) and initiator, 2, 2'-azobis (2-methylpropionamide) dihydrochloride (V-50) were used as received from Sigma, USA. All reactions were performed in deionized water (conductivity 0.055 µS cm$^{-1}$ at 25° C). Acetone was purchased from Rankem, Mumbai, India. Analytical grade sodium nitrate was obtained from Merck, Mumbai, India and used as received.

The homopolymer of 3 acrylamidopropyltrimethyl ammonium chloride (APTMAC) was synthesized by solution polymerization. 2, 2'-azobis(2-methylpropionamide) dihydrochloride (V-50) was used as a thermal initiator. In a typical reaction, 10 g APTMAC (13.3 ml as APTMAC monomer supplied is 75% solution) was dissolved in 90 ml of DI water in a flange type double jacketed reaction vessel with magnetic stir bar and nitrogen gas inlet. The initial total concentration of the monomers in the reaction mixture was 10 wt. %. The reaction mixture was purged with nitrogen gas for 45 minutes to remove any dissolved oxygen. The temperature of the reaction mixture was increased to 56° C. Then, 0.270 g V-50 initiator was added to the reaction mixture with continuous stirring and nitrogen gas purging for 6 h. The viscosity of the reaction mixture increased, indicating the formation of the homopolymer. After completion of the PAPTMAC-Cl polymerization, the polymer was precipitated in acetone, dissolved in water and reprecipitated in (70/30 (v/v)) acetone-water mixture. This procedure was repeated 3 times in order to make sure that the final polymer is free from residual monomer and salt. The chemical structures of the polymers were determined by both $^1$H and $^{13}$C spectroscopy and did not show any extra signals stemming from impurities.

The molecular weight (MW) of the polymer was determined using Agilent 1200 GPC with Shodex OH pak SB-800 series columns. The mobile phase used was 0.30 M $NaNO_3$, with a flow rate of 1.0 ml/min and 50 μL sample injection volume. The GPC column temperature was maintained at 40° C, and polyacrylamide standards were used for calibration. The narrow molecular weight distribution of the polymer (cf. Table 1) is probably due to the V-50 initiator, which has a slow decomposition rate as well as facilitates a monomolecular termination of the growing polymer chains. Furthermore, according to the past information, the monomer APTMAC-Cl inherently contains a chloride anion, which promotes controlled polymerization.[24]

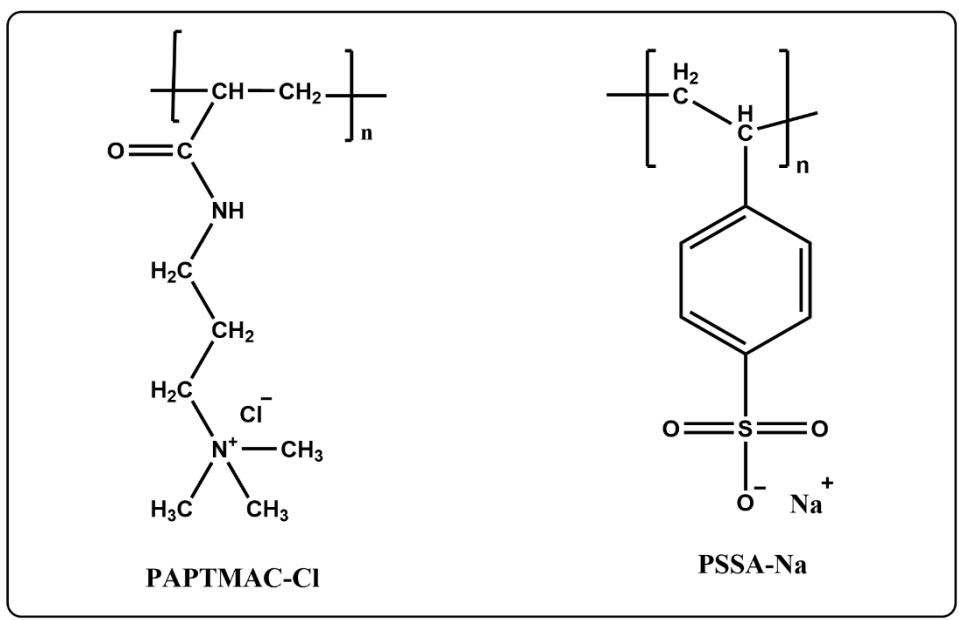

**Scheme 1:** Chemical formulae of the polyelectrolytes under investigation

**PSS-Na 75.6** - poly(styrene sulfonate sodium) - the polyanion under investigation, was prepared from anionically synthesized precursors (characterized by light scattering, viscometry, and GPC) via sulfonation and complete neutralization.[17] The sample under investigation was purchased from Polymer Standard Service (Mainz, Germany) and purified by precipitation and exhaustive dialysis. According to NMR analysis, the degree of substitution is 100%.

**Table 1:** Characteristic data of the polyelectrolytes

| Polymer | $M_w$ kDa | $M_w / M_n$ | $[\eta]$/mL g$^{-1}$ water | $[\eta]$/mL g$^{-1}$ 1 mol/L Na Cl |
|---|---|---|---|---|
| PAPTMAC-Cl | 7.8 | 1.12 | 4 460 | 165.2 |
| PSS-Na | 75.6 | < 1.10 | 1 450 | 13.8 |

**Viscosity measurements** were performed using an Ubbelohde capillary viscometer with capillary diameter 0.63 mm in combination with AVS 470 (SCHOTT Geräte, Mainz, Germany). According to the producer the error in the measured viscosities should remain below 1%. The experiments were carried out within the Newtonian flow regime in pure deionized water (conductivity 0.055 µS cm$^{-1}$ at 25° C) and in NaCl solutions (0.0003 to 1.0 M) at 25° C. The polymer concentrations were varied from 0.005 to 0.075 wt %. Non-Newtonian effects were excluded[17] by measuring the viscosities of the highest molecular weight samples in pure water using three capillaries with different diameters to realize wall shear rates between 70 and 270 s$^{-1}$; the resulting viscosities were the same.

4. Results

**PAPTMAC-Cl 7.8** The analysis of the cluster formation begins with the evaluation of the measured concentration (mass/volume) dependence of the viscosity according to eq (6). Figure 3 show this analysis for solutions of the polycation PAPTMAC-Cl in water and different concentrations of the extra salt, NaCl.

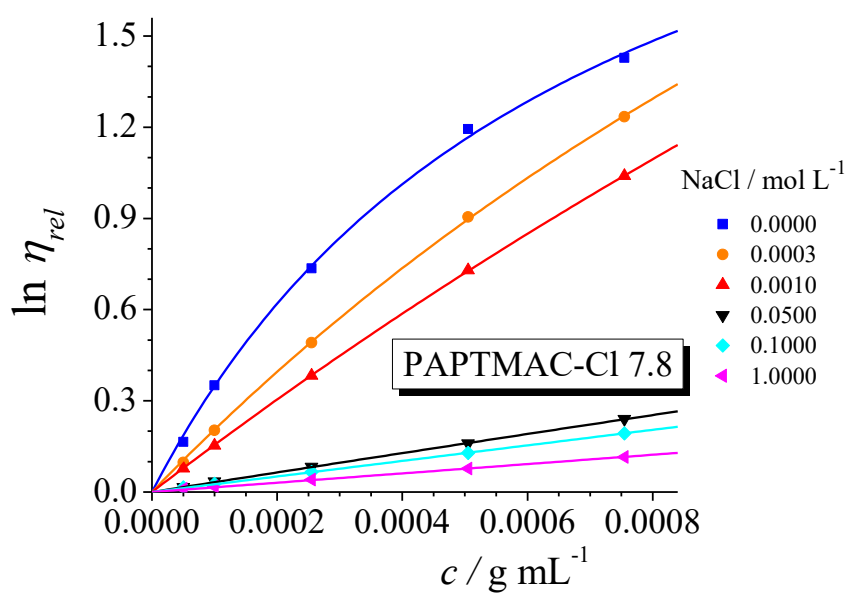

Figure 3. Dependence of the relative viscosities ($\eta_{\text{solution}}/\eta_{\text{solvent}}$) on the concentration of PAPTMAC-Cl in solvents of different salinity. The curves are fits to eq (6).

The parameters obtained from this evaluation are shown in Table 2 for both polyelectrolytes.

The subsequent step of data evaluation consists in the determination of the generalized intrinsic viscosity according to eqs (3) and (8). The results for the present system are shown, in Figure 4. , using the viscometric interaction parameters obtained in the previous step.

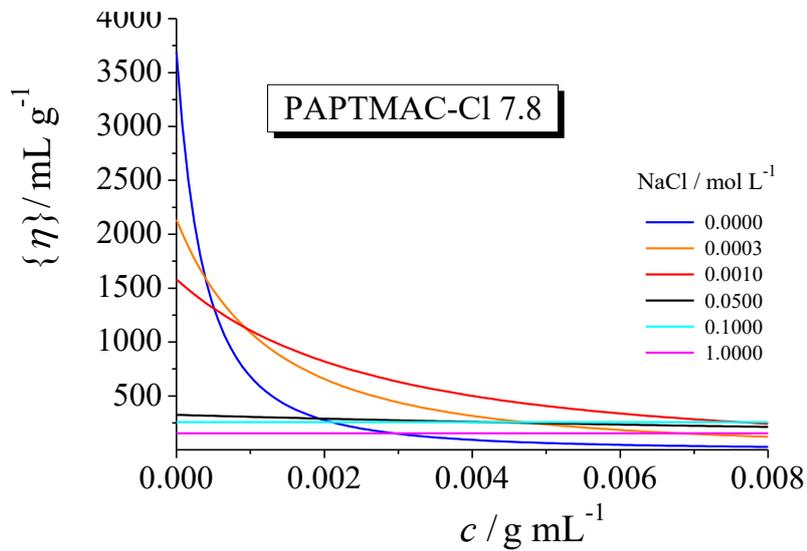

Figure 4. Dependence of the generalized intrinsic viscosities, $\{\eta\}$, on the concentration of PAPTMAC-Cl in solvents of different salinity, calculated according to eq (8) and the parameters of Table 2.as described in Section 2.

Finally, the shear overlap parameter $\Sigma$ of central interest is calculated according to eq (10). Figure 5 displays the results for the present system in the form of a double logarithmic plot for better visualization.

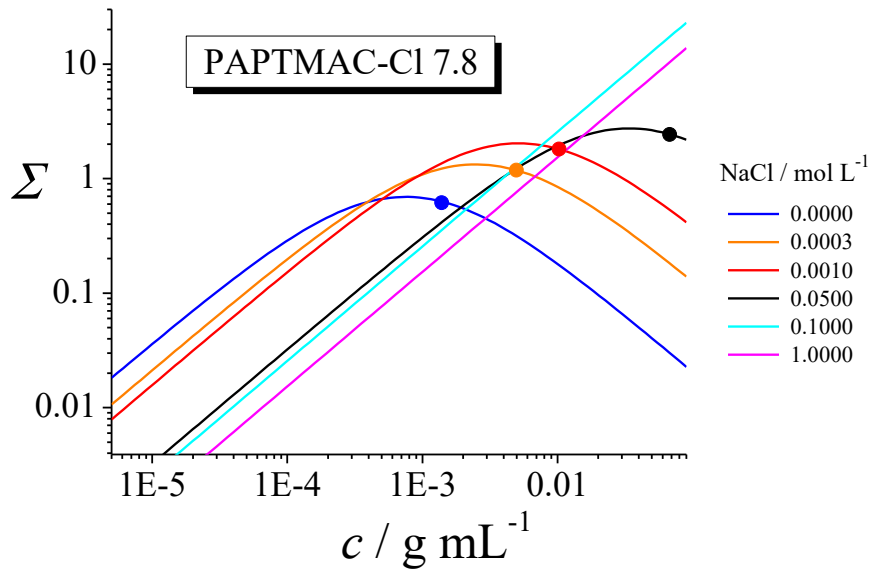

Figure 5. Shear overlap parameter $\Sigma$ as a function of the concentration of PAPTMAC-Cl in solvents of different salinity, calculated according to the eq (10) using the parameters of Table 2. The points represent the cross-over concentrations (cf. eq (11)), separating the solvent-dominated from the polymer-dominated flow regime, which were determined from the inflection points of the double linear plots of the dependence shown in this graph.

**PSS-Na 75.6** For this polyanion, the information concerning the concentration dependence of the relative viscosities in water containing different amounts of NaCl has already been published.[17] The samples were prepared from anionically synthesized precursors by sulfonation and complete neutralization. The obtained polymers were purified by precipitation and exhaustive dialysis. Figure 6 displays the concentration dependence of the generalized intrinsic viscosity.

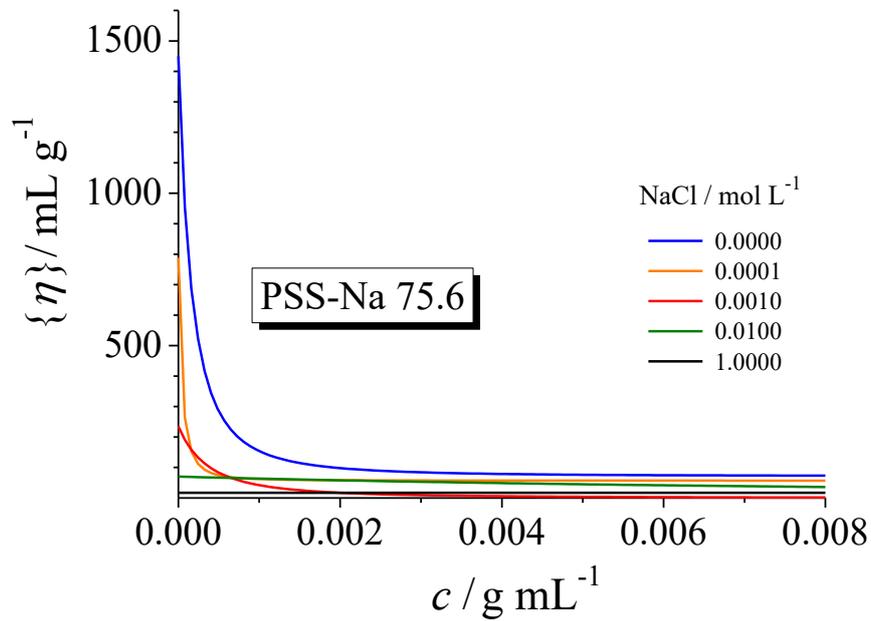

Figure 6. Dependence of the generalized intrinsic viscosities, $\{\eta\}$, on the concentration of PSS-Na in solvents of different salinity, calculated according calculated according to eq (8) and the parameters of Table 2.

Figure 7 displays the corresponding dependence of the shear overlap parameter $\Sigma$.

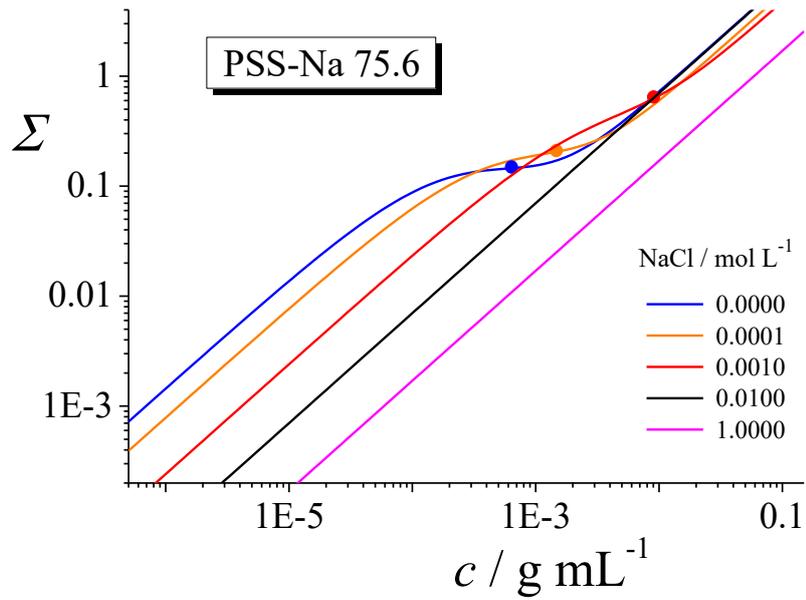

Figure 7. Shear overlap parameter $\Sigma$ as a function of the concentration of PSS-Na 75.6-Na in solvents of different salinity, calculated according to the eq (10) using the parameters of Table 2. The points represent the cross-over concentrations (cf. eq (11)), separating the solvent-dominated from the polymer-dominated flow regime, which were determined from the inflection points of the double linear plots of the dependence shown in this graph.

Table 2: Parameters from the evaluation of ln $\eta_{rel}$ as a function of concentration

| PAPTMAC-Cl 7.8 | | | | | |
|---|---|---|---|---|---|
| Na Cl mol/L | $[\eta]$ g/mL | $\alpha$ | $\beta$ | $c_{crov}$ g/mL | $\Sigma_{crov}$ |
| pure water | 3990 | | 0.361 | 0.0014 | 0.615 |
| 0.0003 | 2140 | | 0.188 | 0.0050 | 1.182 |
| 0.001 | 1580 | | 0.123 | 0.0103 | 1.806 |
| 0.05 | 324 | | 0.091 | 0.0678 | 2.430 |
| 0.1 | 256 | | 0.000 | None | |
| 1.0 | 153 | | 0.000 | None | |
| | | | | | |
| PSS-Na 75.6 | | | | | |
| pure water | 1450 | 0.104 | 2.12 | 0.0006 | 0.146 |
| 0.0001 | 786 | 0.122 | 1.70 | 0.0015 | 0.206 |
| 0.001 | 241 | 0.176 | 0.90 | 0.0136 | 1.224 |
| 0.01 | 70 | 0 | 0.00 | None | |
| 1.0 | 17 | 0 | 0.00 | None | |

## 5. Discussion

The polycation, namely PAPTMAC-Cl and its copolymers with acrylamide finds wide applications in industrial separations, waste water treatments and flocculation.[25-27] The behavior of the polycation PAPTMAC-Cl 7.8 and of the polyanion PSS-Na 75 in solution differs in many ways, where their intrinsic viscosities $[\eta]$ in *pure water* represents the most outstanding dissimilarity, as demonstrated in Figure 8. Despite the fact that the molar mass of PAPTMAC-Cl 7.8 is only about one tenth that of PSS-Na 75, its intrinsic viscosity amounts to $[\eta]_{\text{PAPTMAC-Cl 7.8}}$ = 4460 mL/g, i.e. is much larger than $[\eta]_{\text{PSS-Na 75.6}}$ = 1450 mL/g. This observation is tentatively interpreted in terms of the Pearson HSAB (Hard Soft Acid Base) concept. [12, 13, 28-30]

In the case of PSS-Na 75.6, the sulfonate and the chloride represent hard ions, which means that the ion pair should be stable. The situation for PAPTMAC-Cl 7.8 is different, because the trialkyl ammonium cation is soft as compared to the hard chloride anion, which implies that this ion pair should be considerably less stable. In terms of the degree of dissociation of the ion pairs into free ions, this fact implies that the percentage of free ions on the polymer backbone of PAPTMAC-Cl 7.8 should be considerably larger than in the case of PSS-Na 75.6. The larger intrinsic viscosity of PAPTMAC-Cl 7.8, as compared with that of PSS-Na 75.6, could therefore be caused by the higher electrostatic repulsion in the case of the polycation. The observation that the intrinsic viscosities of the two types of polyelectrolytes become comparable if the solvent contains sufficiently large amounts of extra salt supports this interpretation.

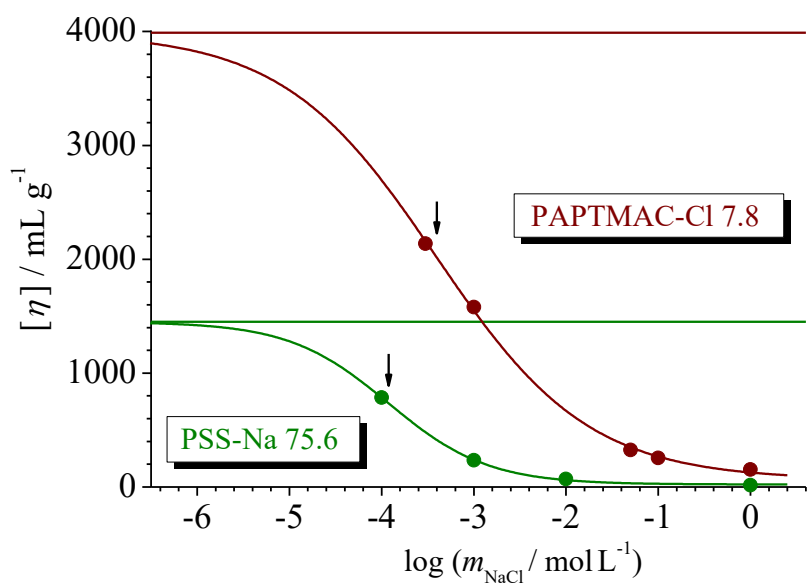

Figure 8. Comparison of the dependence of the intrinsic viscosity $[\eta]$ for the two types of polyelectrolytes on the salinity of the solvent. The arrows indicate the points of inflection. In this and all subsequent Figures in which the behavior of the polyelectrolytes is compared, the curves for the polyanion are shown in brown and those for the polycation in green. The curves are fits to Boltzmann sigmoids.

In order to better understand the completely different concentration dependencies of the coil overlap parameter $\Sigma$ of the two types of polyelectrolytes (Figure 5 and Figure 7), we compare the system-specific generalized intrinsic viscosity $\{\eta\}$ (Figure 4 and Figure 6). In both cases the $\{\eta\}$ values decline monotonously with increasing polymer concentration. The crucial difference lies in an intersection of these curves in the range of low $c$ values in case of the polycation. This dissimilarity leads to the maxima in $\Sigma(c)$ for PAPTMAC-Cl 7.8 and to the monotonous decline for PSS-Na 75.6 shown in Figure 9 for pure water and for the highest salinity.

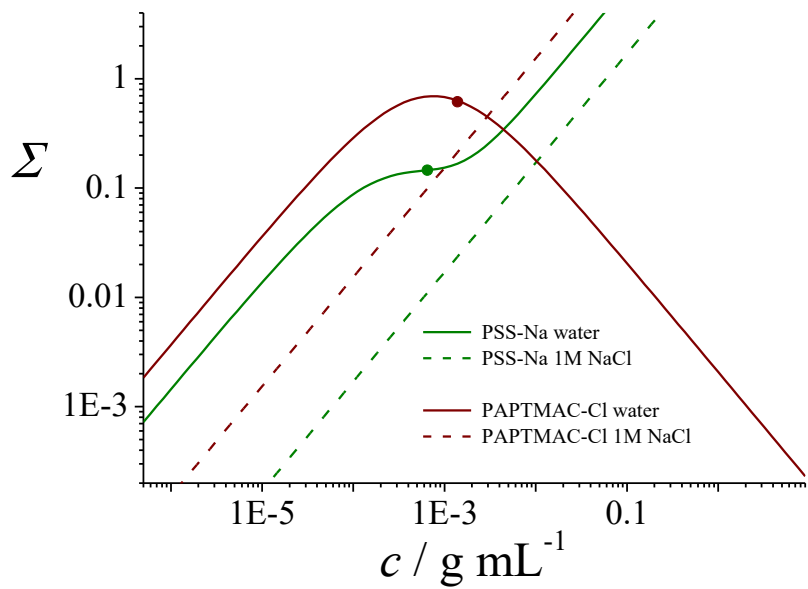

Figure 9. Comparison of the shear overlap parameter $\Sigma$ as a function of the polymer concentration $c$ for solutions of the two types of polyelectrolytes in pure water and in water containing 1 M NaCl, respectively. The curves are calculated according to eqs (10) and (8) using the parameters of Table 2.

One aspect not yet discussed concerns the dependence of the cross-over conditions on the salinity of the solvent and their interrelation. Figure 10 shows $\Sigma_{crov}$ ($n_{NaCl}$), Figure 11 $c_{crov}$ ($n_{NaCl}$) and Figure 12 $\Sigma_{crov}$ ($c_{crov}$).

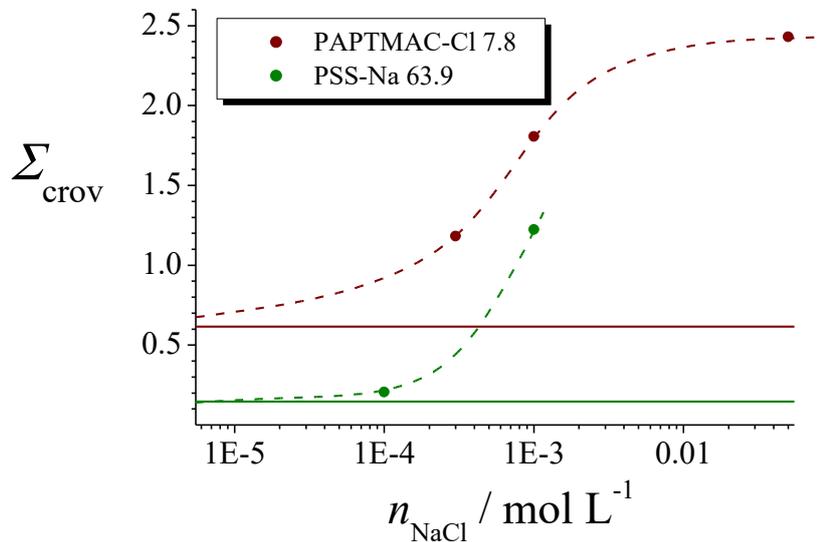

Figure 10. $\Sigma_{crov}$ (shear overlap parameter at the cross-over concentration) as a function of the salinity of the solvent for the two types of polyelectrolytes. The lines parallel to the abscissa represent the asymptotes for pure water. No cross-over phenomena can be observed for the higher NaCl concentrations under investigation. The curves are guides for the eye.

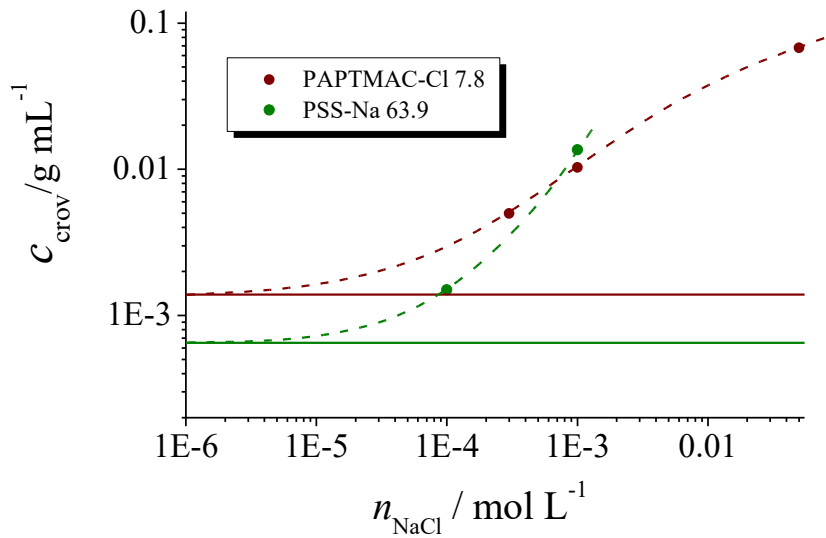

Figure 11. Cross-over concentration, $c_{crov}$, as a function of the salinity of the solvent for the two types of polyelectrolytes. No crossover conditions are observed at the higher salinities under investigation. The curves are guides for the eye.

For pure water and low salinity, the cross-over conditions are larger for PAPTMAC-Cl 7.8 than for PSS-Na 75.6 in accordance with the larger values of the shear overlap parameter. Based on the present results one might speculate that the situation reverses for sufficiently high salinities. At some characteristic concentrations of NaCl the curves end, because $\Sigma_{crov}$ varies linearly with $c$ (cf. Figure 5 and Figure 7) and transitions from solvent dominated to polymer dominated flow behavior can no longer be observed.

The dependence shown in Figure 10 ($\Sigma_{crov}$ vs salt) can be rationalized on the basis of Figure 9. Selecting an arbitrary $c$ in the region of low salinity as a hypothetical $c_{crov}$ value one can see that the corresponding $\Sigma$ is for PAPTMAC-Cl 7.8 always higher than PSS-Na 75.6; again, there is an indication that the situation might reverse at high salt concentrations.

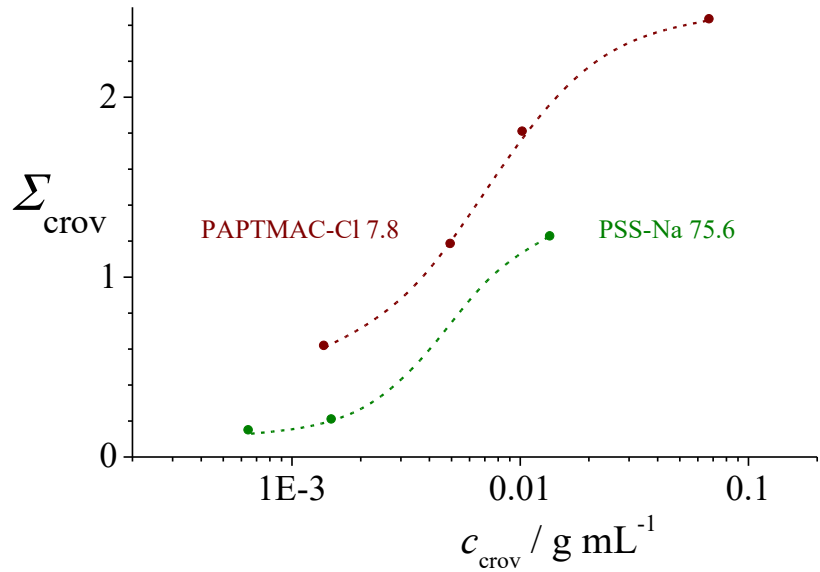

Figure 12. Interrelation of $\Sigma_{crov}$ and $c_{crov}$ for the two types of polyelectrolytes in the solvents of different salinity. The first data point on the two curves (guides for the eye) refers to pure water. For PAPTMAC-Cl 7.8, the salinities of the subsequent data points are $3\cdot10^{-4}$, $3\cdot10^{-3}$ and 0.05 mol/L NaCl, respectively. In the case of PSS-Na 75.6, these values are $10^{-3}$ and $10^{-2}$ mol/L NaCl. No crossover conditions are observed at the higher salinities under investigation. The curves are guides for the eye.

So far, we have only dealt with fully charged linear polyelectrolytes in terms of their hydrodynamic specific volumes and their shear overlap parameters. This information is essential for the theoretical understanding. For practical purposes, however, it is probably more interesting to extend the studies to partly charged polymers, non-linear polyelectrolytes and to polyelectrolytes with block structure to name but a few.

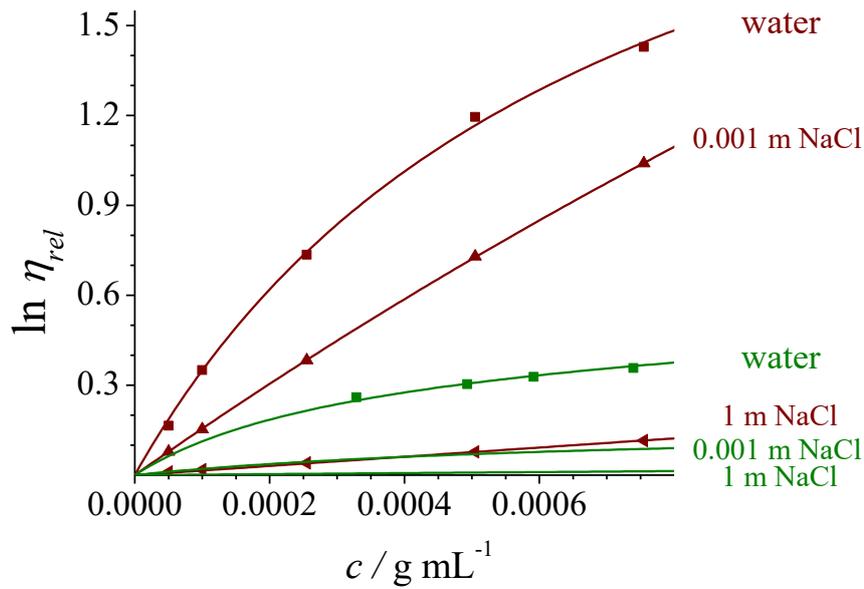

Figure 13. Comparison of the concentration dependence of ln $\eta_{rel}$ for solutions of PAPTMAC-Cl 7.8 (brown) and of PSS-Na 75.6 (green) in solvents of the same salinity.

Figure 13 shows ln $\eta_{rel}$ as a function of concentration for pure water and for salinities of 0.001 and 1.00 molar NaCl, respectively. Despite the almost ten times lower molar mass of PAPTMAC-Cl 7.8 the viscosity of its solution in water is for instance at 0.6 g/L by a factor of 2.6 times larger than for PSS-Na 75.6. This ratio still amounts to 2.2 for 0.001 mol/L and even for 1 mol/L it is larger than unity, namely approx. 1.09. This observation clearly demonstrates how decisive it is for the viscosities of the solutions whether the ions of the polyelectrolytes are of the hard/weak type (PAPTMAC-Cl 7.8) of the hard/hard type (PSS-Na 75.6). This feature should open up new possibilities to tailor desired viscometric behaviors polyelectrolyte solutions.

Up to this point all considerations were purely phenomenological. In order to gain some first insight concerning the molecular background of the observed phenomena, we describe ln $\eta_{rel}$ as a series in $\tilde{c}$ as formulated in eq (7). The results calculated from the data of Table 1 are collected in Table 3

Table 3: Second and third coefficient of the

series expansion of ln $\eta_{rel}$ (eq (7))

|  | NaCl mol/L | $\alpha - \beta$ | $\beta^2 - \alpha\beta - \gamma$ |
|---|---|---|---|
| PAPTMAC-Cl 7.8 | 0.000 | -0.36 | 0.13 |
|  | 0.001 | -0.12 | 0.02 |
|  | 1.000 | 0.00 | 0.00 |
| PSS-Na 75.6-Na 75.6 | 0.000 | -2.02 | 4.27 |
|  | 0.001 | -0.72 | 0.65 |
|  | 1.000 | 0.00 | 0.00 |

The signs of the coefficient are equal for both types of polyelectrolytes. This means that binary contacts between the solute **de**crease the viscosity and ternary contacts **in**creases it. There is, however, a significant difference in the absolute values for PAPTMAC-Cl 7.8 and for PSS-Na 75.6: For pure water and low salinities they are always considerably lower in the former than in the latter case. These differences disappear as the salt concentration in the solvent surpasses a characteristic value. Under these conditions, the electrostatic shielding is so strong that the viscosity as a function of composition is in the concentration range under investigation exclusively determined by $M_1$, i.e. by the intrinsic viscosity of the polyelectrolyte.

### 6. Conclusions/Outlook

This study was initiated to find out whether there exist fundamental differences in the shear clustering behavior of polyanions and polycations. Indeed, one does find many dissimilarities for the two types of polyelectrolytes. The *intrinsic viscosities* $[\eta]$ of PAPTMAC-Cl 7.8 in water result three times higher than that of PSS-Na 75.6, despite the fact that its molar mass is only approximately one tenth. This peculiarity persists for the *generalized intrinsic viscosities*, (Figure 4 and Figure 6), where the concentration dependence of $\{\eta\}$ is again much higher for the polycation that for the polyanion.

In the region of low solvent salinities, the described situation leads to shear overlap parameters $\Sigma$ which differ fundamentally in their composition dependence: For PAPTMAC-Cl 7.8 $\Sigma(c)$ exhibits a maximum in contrast to PSS-Na 75.6, which displays a point of inflection. For sufficiently high salt contents of the solvent, the composition dependence of $\Sigma$ becomes linear in both cases.

For uncharged products and for solutions of polyelectrolytes in solvents of low salinities, one always observes points of inflection (cross-over points) as the solute concentration rises; they mark the transition from solvent to solute dominated flow behavior. In plots of $\Sigma_{crov}$ or $c_{crov}$ *vs* the salinity of the solvent and in plots of $\Sigma_{crov}$ *vs* $c_{cross-over}$, the curves for PAPTMAC-Cl 7.8 are always situated above those for PSS-Na 75.6 in agreement with the higher $[\eta]$ and $\{\eta\}$ values of the polycation.

In the absence of additional information, it is impossible to decide, whether the observed fundamental differences in the behavior of polycations and polyanions result from the opposite sign of the charges fixed on the polymer backbone or whether they are caused by other factors. On a trial basis we have therefore investigated whether the HSAB concept of Pearson could explain the observed differences and found out that this approach is indeed in the position to make all experimental observations comprehensible. The reason for the dissimilarities is the unequal combination of hard and soft ions for the two types of polyelectrolytes. With PAPTMAC-Cl 7.8 the soft $NR_4^+$ ion goes along with the hard $Cl^-$ counterion and with PSS-Na 75.6 the hard $RSO_3^-$ is combined with the hard counterion $Na^+$. This situation implies that the degree of charging of the polymer backbone results much higher in the former case than in the latter case with the consequence that the $[\eta]$ and $\{\eta\}$ values of the polycation are much larger than that of the polyanion despite its low molar mass. This fact suffices to explain all the observed differences in the behavior of the two types of polyelectrolytes.

The behavior patterns just described lead directly to very different viscosities of the solutions of the two types of polyelectrolytes at the same concentrations and salinities. In all cases the solutions of PAPTMAC-Cl 7.8 are much higher than that of PSS-Na 75.6; the effects increase with rising *c* and are largest for pure water, where they assume almost three times higher values. Even for a large excess of extra salt the increase is still noticeable.

In order to obtain a certain minimum of molecular information in addition to the results of the above phenomenological discussion, we have developed the measured $\ln \eta_{rel}$ values into a concentration series and determined their coefficients. This treatment reveals that the sign of these coefficients is the same for both types of polyelectrolytes; however, the absolute values are very different: The formation of binary intermolecular contacts between the solute *reduces* the viscosity of for PAPTMAC-Cl 7.8 in pure water and solvents of low salinity much less than for PSS-Na 75.6. The formation of ternary contact on the other hand *increases* the viscosity of the PAPTMAC-Cl 7.8 solutions much less than that of the PSS-Na 75.6 solutions.

Building on the extensive and well-established research demonstrating the broad applicability of the HSAB concept for low molecular weight substances [12, 13, 29], it seems reasonable to extend this principle to charged macromolecules as well. Current understanding of polyelectrolytes, however, remains limited and largely confined to linear homopolymers, while only a handful of studies have explored charged copolymers[31], non-linear systems[7], or biopolymers.[32]

The authors believe that the present theoretical framework offers a promising pathway: by using HSAB-guided counterion selection, it may become possible to design polyelectrolytes with tailored properties. Such an approach could prove transformative in fields where polymers already play a vital role, including electronics, medicine, self-healing materials, and advanced manufacturing.

## 7. Associated content

None

## 8. Author information

**Corresponding author**

**Bernhard A. Wolf** - *Johannes Gutenberg-Universität Mainz, Department Chemie, Jakob Welder-Weg 11, D-55099 Universität Mainz, Germany;*

orcid.org/0000-0002-4051-7114

E-mail: bernhard.wolf@uni-mainz.de

**Authors**

**Manohar V. Badiger** *Polymer Science and Engineering Division, National Chemical Laboratory (NCL), Pune- 411 008, orcid.org/*0000-0001-9265

E-mail: mv.badiger@ncl.res.in

**Suresha P. Ranganath** *Polymer Science and Engineering Division, National Chemical Laboratory (NCL), Pune- 411 008,* orcid.org/0000-0001-6703-3585

E-mail: pr.suresha@ncl.res.in

**ACKNOWLEDGEMENTS**

MVB thanks the Alexander von Humboldt (AvH) Foundation Bonn, Germany for a revisit fellowship.

Furthermore, the authors would like to thank Prof. W. Tremel (University of Mainz, Germany) for pointing out the potential of R.G. Pearson's HSAB concept for interpreting the experimental results.